\documentclass{emulateapj}

\usepackage{epsfig}
\usepackage[]{natbib}

\newcommand{\eg}{{\rm e.g.,}}
\newcommand{\ie}{{\rm i.e.,}}

\newcommand{\Lo}{\ensuremath{L_\odot}}

\newcommand{\asec}{\ensuremath{\arcsec}}

\shorttitle{AEGIS: IRAC-selected EROs}
\shortauthors{Wilson et al.}

\begin{document}

\notetoeditor{This letter is intended to appear in the AEGIS special issue}

\title{AEGIS: A PANCHROMATIC STUDY OF IRAC-SELECTED EXTREMELY RED OBJECTS WITH CONFIRMED SPECTROSCOPIC REDSHIFTS}

\author{G. Wilson\altaffilmark{1}, J.-S. Huang\altaffilmark{2}, G.~G. Fazio\altaffilmark{2},  R. Yan\altaffilmark{3}, A. M. Koekemoer\altaffilmark{4}, S. Salim\altaffilmark{5}, S.~M. Faber\altaffilmark{6}, J. Lotz\altaffilmark{6}, C.~N.~A. Willmer\altaffilmark{7}, M. Davis\altaffilmark{3}, A.~L. Coil\altaffilmark{7,8}, J.~A. Newman\altaffilmark{8,9},  
 C.~J. Conselice\altaffilmark{10},  C. Papovich\altaffilmark{7}, M.~L.~N. Ashby\altaffilmark{2}, P. Barmby\altaffilmark{2}, S.~P.~Willner\altaffilmark{2}, R. Ivison\altaffilmark{11}, S. Miyazaki\altaffilmark{12}, and D. Rigopoulou\altaffilmark{13}
}

\altaffiltext{1}{Spitzer Science Center, California Institute of Technology, 220-6, Pasadena, CA 91125; gillian@ipac.caltech.edu}

\altaffiltext{2}{Harvard-Smithsonian Center for Astrophysics, 60 Garden Street, Cambridge, MA 02138}

\altaffiltext{3}{Department of Astronomy, University of California Berkeley, Campbell Hall, Berkeley, CA94720}

\altaffiltext{4}{Space Telescope Science Institute, 3700 San Martin Drive, Baltimore, MD 21218}

\altaffiltext{5}{Department of Physics and Astronomy, University of California Los Angeles, Knudsen Hall, Los Angeles, CA 90095}

\altaffiltext{6}{UCO/Lick Observatory and Department of Astronomy and Astrophysics, University of California Santa Cruz, 1156 High Street, Santa Cruz, CA 95064}

\altaffiltext{7}{Steward Observatory, University of Arizona, Tucson, AZ 85721}

\altaffiltext{8}{Hubble Fellow}

\altaffiltext{9}{Institute for Nuclear and Particle Astrophysics, Lawrence Berkeley National Laboratory,Berkeley, CA 94700}

\altaffiltext{10}{The School of Physics and Astronomy, University of Nottingham, University Park, Nottingham NG7 2RD, UK}


\altaffiltext{11}{Astronomy Technology Centre, Royal Observatory, Blackford Hill, Edinburgh EH9 3HJ, U.K.}

\altaffiltext{12}{Subaru Telescope, National Astronomical Observatory of Japan, 650 North A'ohoku Place, Hilo, HI 96720}

\altaffiltext{13}{Department of Astrophysics, Oxford University, Keble Road, Oxford, OX1 3RH, United Kingdom}

\begin{abstract}

We study 87 Extremely Red Objects (EROs), selected both
to have color redder than~$R-[3.6]=4.0$, and to have confirmed spectroscopic
redshifts. Together, these two constraints result in this sample populating a fairly narrow redshift range at $0.76 < z < 1.42$.
The key new ingredient included here is deep \emph{Spitzer} Space Telescope InfraRed Array Camera (IRAC) data.
Based on $[3.6]-[8.0]$ color, we demonstrate that it is possible
to classify EROs into early-type, dusty starburst, or power-law (AGN) types.
We present ultraviolet to 
mid-infrared spectral energy distributions (SEDs) and Advanced Camera for Surveys (ACS) images, both of which support 
our simple IRAC color classification.

\end{abstract}


\section{INTRODUCTION}
\label{sec:intro}

First discovered in the late 1980s 
\citep{elston-88},
extremely red objects (EROs) are 
defined by their very red optical/near-infrared colors.
It has been known for some time that the redness of their color constrains these galaxies to 
be either early-type 
galaxies, starburst galaxies reddened by dust, 
or AGN  (or a combination of these three classes).
However, until recently, with observations limited to $K$-band (or shorter wavelengths),
it has proven extremely challenging to accurately classify the EROs by type \citep{mann-02}, even
in combination with high-resolution HST imaging \citep{mous-04}.

Although EROs appear to consist of a heterogeneous mix of galaxy classes, the emerging paradigm
is that they may well be the high redshift counterparts and progenitors of local massive E and SO galaxies.
Their reliable classification and study, especially at intermediate redshift ($z \sim 1$) can provide crucial constraints on the evolution of massive, starburst,  dusty and/or ultraluminous infrared galaxies (ULIRGs) known to exist at higher redshift \eg\ BzKs \citep{daddi-04}, BX/BMs \citep{reddy-05}, Distant Red Galaxies  \citep[DRGs]{franx-03, pap-06a, conselice-06}, submillimeter and IR-Luminous Lyman Break Galaxies \citep[ILLBGs]{huang-05, rig-06}.


In the restframe near-IR, old stellar populations
show a turndown at wavelengths longer than  the restframe $1.6 \micron$ `bump', while dusty starburst populations show emission from small hot dust grains.
AGN-dominated sources display a power-law spectral energy distribution.
In \cite{wilson-04} we showed how data from \emph{Spitzer} could  begin  to help 
distinguish among different ERO populations. 
In this paper, we extend our ERO study to take advantage of the rich panchromatic 
dataset available from the
All-wavelength Extended Groth strip International Survey (AEGIS).
All magnitudes used in this letter are AB, unless otherwise specified. 


\section{THE AEGIS DATASET}
\label{sec:obs}

The \emph{Spitzer} IRAC ($3.6,4.5,5.7,8.0\micron$, \citealt{fazio-04})
component of the Extended Groth Strip (EGS) survey 
spans an area of $120 \times10$ arcmin (Huang et al. 2006, in prep: see also \citealt{huang-04, barmby-06, huang-06b}). 
In conducting this ERO study, 
we also utilized DEEP2 spectroscopy (Davis et al. 2006, in prep), and
$u^{\prime} g^{\prime}$ (Ashby et al. 2006, in prep), CFHT
$BRI$ \citep{coil-04a}, ACS $V(F606W)$ and $I(F814W)$, deep Subaru $R$ \citep[27.0 AB,  $5\sigma$]{miya-02}, $K$ (Conselice et al. 2006, in prep), and \emph{Spitzer} Multiband Imaging Photometer (MIPS, \citealt{rieke-04}) $24 \micron$ imaging. Further details may be found in \citet {davis-06}.

\section{IRAC-SELECTED GALAXIES AND THEIR COLOR-COLOR DISTRIBUTION}
\label{sec:color-color}


The EGS field contains $\sim 45000$ galaxies detected at $3.6 \micron$ (23.9 AB, $5\sigma$).
Figure~\ref{fig:R14} shows an $R-[3.6]$ versus $[3.6]-[8.0]$ 
color-color diagram. The black points show the $\sim 13000$  $3.6\micron$-selected galaxies
with both good quality  Subaru $R$ and  $8.0 \micron$ photometry. 

The colored tracks on Figure~\ref{fig:R14} show the location in color-color space as a function of redshift ($0 < z< 4$) for eight non-evolving empirical templates (four common \citealt*{cww-80} (CWW) templates [E, Sbc, Scd, Im] empirically extended to $10 \micron$ using ISO data (Huang et al., 2006, in prep),
a dusty starburst template [M82], a dusty starburst/ULIRG template [Arp220], and two AGN templates [NGC 1068 and NGC 5506]). 
While we fully expect these templates to become increasingly inaccurate at high redshift, they do serve
to provide simple insight onto the likely nature and redshift distribution of galaxies 
within this color-color diagram. 

A distinctive swath of galaxies is clearly apparent curving from the left to the top of Figure~\ref{fig:R14} (the plume in the lower left corner is caused by stellar contamination).
Notice, especially, the excellent agreement between this swath of galaxies observed in color-color space and the CWW E (blue) track to $z \sim 0.7$, when the template noticeably begins to diverge from the ``bluer'' data.  
We note that the predictability of the $[R] -[3.6]$ color-redshift relation 
for early-type galaxies can be utilized as an effective technique 
for detecting high redshift clusters of galaxies \eg\ 
at $z < 1.4$ in the $4 \deg^{2}$
Spitzer First Look Survey Field ({\citealt{wilson-05}; Muzzin et al, 2006, in prep}). 
The SpARCS collaboration\footnote{http://spider.ipac.caltech.edu/staff/gillian/SpARCS}
is currently utilizing an even redder $[z^{\prime}] -[3.6]$ color to detect and study clusters to $z=2 $ in the $50 \deg^{2}$ Spitzer SWIRE Legacy Fields
\citep{wilson-06a}.

We \emph{define} an ERO to be a galaxy redder in color than $R-[3.6] = 4.0$. This is the same criterion used in \cite{wilson-04}, and is
a very similar selection criterion to 
the traditional Vega  $R-K >5.0$ requirement (see \citealt{wilson-04} for a discussion). 
 
From Figure~\ref{fig:R14}, we might expect that only high redshift ($z\ga 0.8$) early-type, ($z\ga 1.0$) dusty starburst galaxies, and AGN (at any redshift) would satisfy this extremely red criterion. 
As we shall demonstrate in the remainder of this letter, this indeed turns out to be the case.
Note that late-type CWW Scd (cyan) and irregular (purple) galaxies are \emph{never}
sufficiently red to be classified as an ERO \emph{at any redshift}, and one would not expect
to find any CWW late-type Sbc (green) EROs at $z < 1.5$.

\section{REDSHIFT DISTRIBUTION OF THE SPECTROSCOPIC ERO SAMPLE}
\label{sec:zdist}


There are several thousand EROs in the EGS field.
Here, we carry out a pilot study of the 87 EROs 
with confirmed DEEP2 spectroscopic redshifts.
We use spectroscopy in this letter only for redshift determination.

The solid black histogram in the far left panel of Figure~\ref{fig:hist} shows the redshift distribution of the 87 EROs in our sample. The EROs occupy a relatively  narrow redshift range at $0.76  < z < 1.42$. The dotted black histogram shows the  redshift distribution of those galaxies in
the DEEP2 EGS field 
with good quality spectroscopic redshift determinations (scaled down by a factor of 15).

The fact that none of the EROs are located at $z <0.76$ is not a DEEP2 selection effect; 
in EGS, DEEP2 samples the full redshift range $0 < z < 1.45$ \citep{faber-06, willmer-06}. 
It is  caused by the additional ERO $R-[3.6]> 4.0$ selection requirement, which effectively
excludes low redshift galaxies (Figure~\ref{fig:R14}).
The DEEP2+ERO selection function also limits this sample to a fairly bright but narrow range of 
$18.5< [3.6]< 20$, with the one exception of 12007954, a 17th magnitude AGN \citep{lefloch-06}.
Note that by selecting only those EROs with spectroscopic redshifts, we may bias our sample 
against inclusion of any low-redshift extremely dusty galaxies
(since optically faint galaxies will not pass the 
$R<24.1$ selection requirement
of the DEEP2 survey).

\section{ERO Classification Using [3.6]-[8.0] Color}
\label{sec:IRACcolorclass}

It is possible to divide the EROs in this sample into three broad classes by means of
their IRAC [3.6]- [8.0] color.
In this section, we introduce each of these classes, show examples of their SEDs and
discuss the sometime subtle features which are only apparent from such broad-baseline
datasets. We also study their morphologies using ACS images.

From Figure~\ref{fig:R14}, early-type galaxies are expected
to have the bluest  ($[3.6]- [8.0] \simeq -1.0$) colors, dusty starburst galaxies 
intermediate colors  ($[3.6]- [8.0] \simeq -0.5$), and AGN reddest colors ($[3.6]- [8.0] \simeq +1.0$).
The right panel of Figure~\ref{fig:hist} corresponds to the dashed rectangle in Figure~\ref{fig:R14} and shows the subsample of 87 EROs with spectroscopic redshifts. 
We \emph{classify} $[3.6]- [8.0] < -0.75$ galaxies as bulge-dominated early types (blue circles), $- 0.75< [3.6] - [8.0]> 0.0 $ galaxies as dusty starburst\footnote{We use the word ``starburst'' to mean
a galaxy whose IR luminosity is powered by star formation. 
We do not intend to imply these galaxies have total  IR luminosity $L(8-1000 $\micron$) < 10^{11} \Lo$. Indeed, as we shall discuss in \S~\ref{sec:disc} most of these dusty starburst EROs would actually be classified as LIRGs.} (red triangles) and [$3.6] - [8.0]< 0.0$ galaxies as 
power-law AGN types (orange boxes). 
As we shall demonstrate, these two simple color cuts do separate the classes 
rather well. This is because the redshift distribution of this particular sample is rather narrow. 
For a sample spanning a broader redshift distribution, one might imagine utilizing a more 
complex color selection to isolate the meanderings of the different populations 
through color-color space in Figure~\ref{fig:R14}.

We classify 53 of the EROs as early types.
They are located at $0.82  < z < 1.28 $ and their redshift distribution is shown by the blue hatching in the middle panel of Figure~\ref{fig:hist}. We classify 32 EROs as being dusty starburst types. The dusty starburst population (shown by the red hatching) has a slightly wider redshift 
distribution from $0.76  < z < 1.42 $ (presumably because galaxies in this class 
can contain arbitrary amounts of dust).
We classify two EROs as being predominantly power-law types.  
These are 12007878  at $z = 0.99$ and 12007954 at $z = 1.15$, and are shown in solid orange 
(see also \citealt{konidaris-06}).

\section{Example Spectral Energy Distributions and ACS Images}
\label{sec:seds}

Figures~\ref{fig:seds_E} and~\ref{fig:seds_DSB} (Plate 1)
show examples of observed SEDs and 
$3 \arcsec \times 3 \arcsec$  $I$-band (F814W) postage stamp ACS images. 
For comparison, also shown are CWW E (blue) and M82 (red) templates (\S~\ref{sec:color-color}) 
as they would appear at the redshift of each ERO.
At $z \sim1$, IRAC's  $8 \micron$ channel
measures any emission from small hot dust grains (indicative of a dusty starburst galaxy) 
while  the $3.6 \micron$ channel measures the stellar peak. A relative red (or blue) $[3.6]-[8.0]$ color, therefore, can be used to discriminate for (or against) dusty starburst galaxies (and AGN) at this redshift.

\subsection{Early-Type EROs}
\label{ssec:E}

Figure~\ref{fig:seds_E} (Plate 1) shows 
six examples of galaxies IRAC color-classified as early-type EROs. 
The CWW E template approximates the 13048898, 13019047, 13004276, and
12004426  SEDs very well. The CWW E template appears to underestimate the SEDs of
13019309 and 12008091 in the UV.
In all cases the ACS images show clear 
evidence of bulge-dominated  morphologies, supporting our color classification as early types. 

A total of 35 of the 53 early-type EROs fall within the ACS footprint.
A close inspection of all of the available ACS images reveals them to be predominantly bulge-dominated 
(E, S0 or Sa-type spirals). A small number ($\sim 10 \%$) appear to be undergoing mergers.
We conclude that choosing a blue [3.6]-[8.0] color successfully selects for $z=1$ EROs with old
stellar populations and against those with dusty starburst or power-law features.

\subsection{Dusty Starburst and Power-Law EROs}
\label{ssec:M82}

Figure~\ref{fig:seds_DSB}  (Plate 1) shows five examples of galaxies IRAC color-classified
as dusty starburst EROs, and one 
power-law ERO (12007878).
The M82 template approximates the SEDs of sources 13042940, 12100899, and  12007831 very  well,
and that of 13004291 reasonably well.
In addition to being more mid-IR luminous than an early-type ERO of similar $3.6\micron$ magnitude, 
the SEDs of these four sources are more luminous in the optical.
Their ACS images show clear evidence of disturbed, peculiar, interacting or merging galaxies.

Neither template approximates well the SED of 12008048. Close inspection of the ACS image
reveals it to be a face-on spiral galaxy. Since a normal spiral at $z = 0.9$ 
would not meet the $R-[3.6]$
redness criterion, we infer that 12008048 must be an especially dusty spiral.

A total of 17 of the 32 EROs we classify as dusty starbursts fall within the ACS footprint.
Close inspection of all of their images reveals many of them ($\sim 60\%$) to
be disturbed or interacting galaxies, and the remainder ($\sim 40\%$)  to be 
late-type spirals.

12007878 shows a  monotonically increasing power-law SED. Its host galaxy is clearly 
bulge-dominated and it can be unequivocally classified as an AGN-dominated source.



\section{DISCUSSION}
\label{sec:disc}



Turning to those EROs which have $24 \micron $ detections, there are five ($17.4 < [24] < 19.1$) amongst the 53 early-type
and 29 ($16.8 < [24] < 19.1$) amongst the 
32 dusty starburst EROs. Both power-law EROs have a $24\micron$ detection (14.5 and 16.5). 
In four out of the five cases of early-type EROs with a $24 \micron $ detection, 
an ACS image is available.
In all four cases, a merging galaxy with several
close neighbors is apparent. The MIPS point-spread function is $6 \asec$ so it is possible that the signal in each case might be associated with another galaxy entirely.

We use  \citet[Fig.~7]{lefloch-05} to translate our $24\micron$ flux limit 
(80uJy, 19.1 AB, $5\sigma$) into a lower limit on the total detectable IR luminosity 
of an ERO in our redshift range. Depending on the model used, at $z \sim 0.75$, we are capable of
detecting
sources more luminous than $\sim 5 \times 10^{10} \Lo$ \ie\ some dusty starbursts, and all LIRGS
and ULIRGS. At $z \sim 1.4$, we are capable of detecting
sources more luminous than $\sim 5 \times 10^{11} \Lo$ \ie\ LIRGS
and ULIRGS. Based on their actual $24\micron$ magnitudes and redshifts, most of
our ``starburst'' sources have intrinsic luminosities
between $10^{11}$ and $10^{12}$ $\Lo$, \ie\ LIRG class. Both
of the ``AGN'' sources have intrinsic luminosities of $ > 10^{12} \Lo$,
putting them in the  ULIRG class.

Although, in this study, we utilized optical/IRAC colors to classify our ERO sample, IRAC-[24] color selection may be a more effective discriminator at higher redshift \citep{lacy-04, ivison-04, sajina-05}.

Putting IRAC ERO colors in the context of perhaps more familiar restframe  optical colors,
Figure~\ref{fig:UB14}  shows \emph{restframe} $U-B$ color versus \emph{observed} $[3.6]-[8.0]$ color for all IRAC-detected galaxies with
good [8.0] and good CFHT BRI photometry (see Willmer et al. 2006
for details of the conversion to restframe $U-B$). 
The symbols are as in Figure~\ref{fig:hist}. The filled circles in Figure~\ref{fig:UB14} denote the 36 EROs which are also detected at $24\micron$ (see also Figure~\ref{fig:hist}). 
The EROs we classify as early types generally tend  to lie on the $U-B$ defined ``red-sequence'' and  the 
 EROs we classify as dusty starburst  tend to lie in the ``green valley'' and at the top
of the ``blue cloud'' \citep{weiner-05}. 

\section{CONCLUSIONS}
\label{sec:conc}

In this paper, we utilized the AEGIS dataset
to explore the nature of 87 EROs with
confirmed spectroscopic redshifts. 
IRAC imaging facilitated 
dividing this sample into three distinct classes using a simple [3.6] -[8.0] 
color selection technique.
We presented SEDs and high-resolution ACS images supporting our (early-type, 
dusty starburst or power-law) classification.

\begin{figure}
\plotone{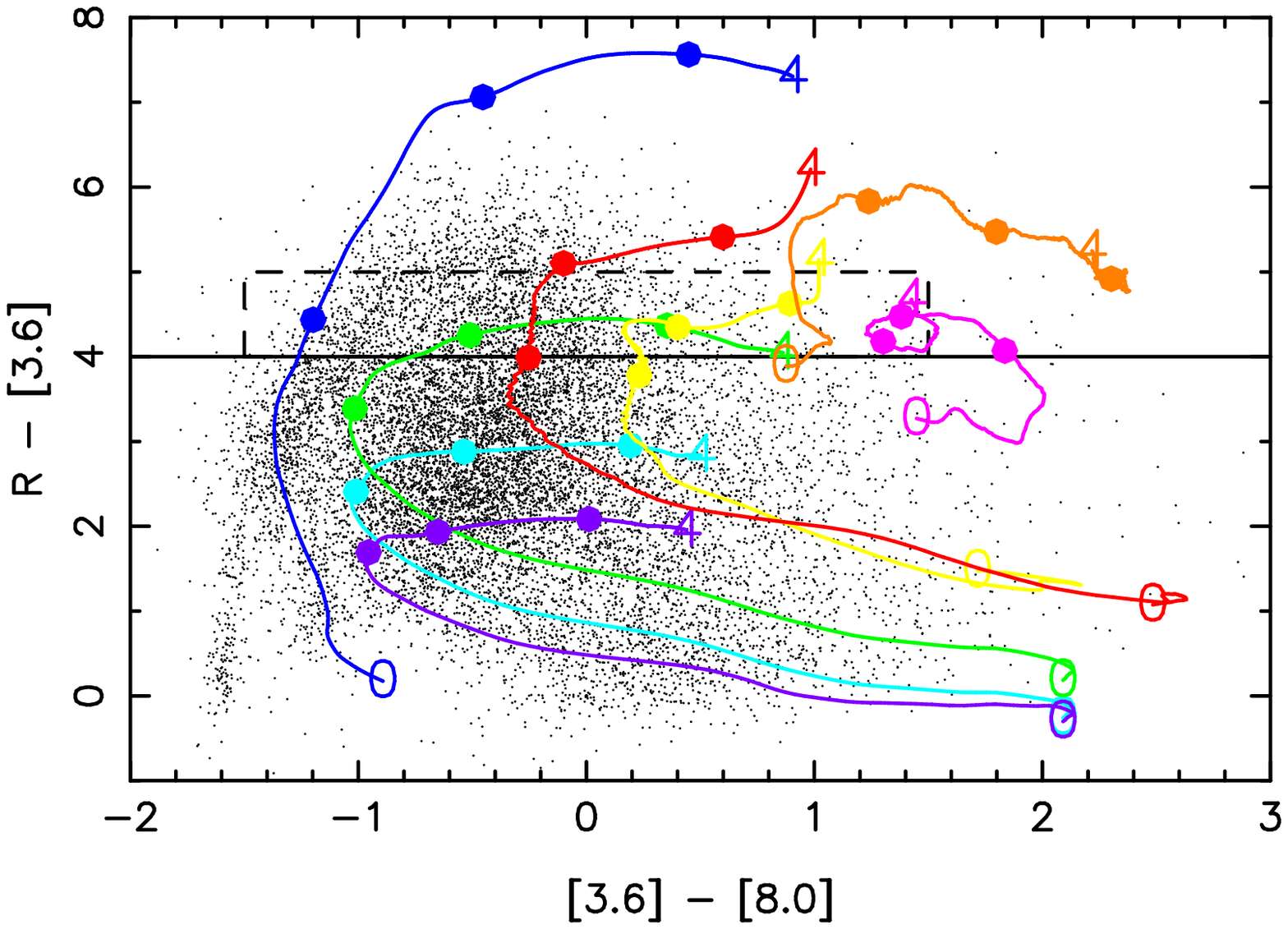}
\caption{
$R-[3.6]$ versus $[3.6]-[8.0]$ color-color diagram for IRAC-selected galaxies. 
The solid lines show non-evolving templates in color-color space
for CWW early-type (blue),  CWW Sbc (green), CWW Scd (cyan), CWW Im (purple), dusty starburst M82 (red), dusty starburst/ULIRG Arp220 (yellow),
NGC 1068 (pink) and NGC 5506 (orange).
The redshift is indicated for each type. 
A distinctive swath of galaxies is apparent curving from the left to the top of the Figure. 
The CWW early-type (blue) track follows this  extremely well to $z \sim 0.7$, when the 
template noticeably begins to diverge from the ``bluer'' data.  
The plume in the lower left corner is caused by stellar contamination.
Galaxies with $R-[3.6]> 4.0$ (\ie\ above the black line) are defined as EROs. 
One would expect only high redshift ($z > 0.8$) early-type, ($z >  1.0$) dusty starburst galaxies, ($z >  1.5 $) CWW Sbc, and 
AGN (at any redshift) to satisfy this extremely red criterion. The dashed black rectangle 
indicates the region of color-color space populated by the
87 EROs with confirmed spectroscopic redshifts (see also right panel of Figure~\ref{fig:hist} and \S~\ref{sec:IRACcolorclass}).
\label{fig:R14}
}
\notetoeditor{I do not know how to convert any of my figures
from RGB to CMYK. I will need help with this.}
\end{figure}

\begin{figure}
\epsscale{.50}
\plotone{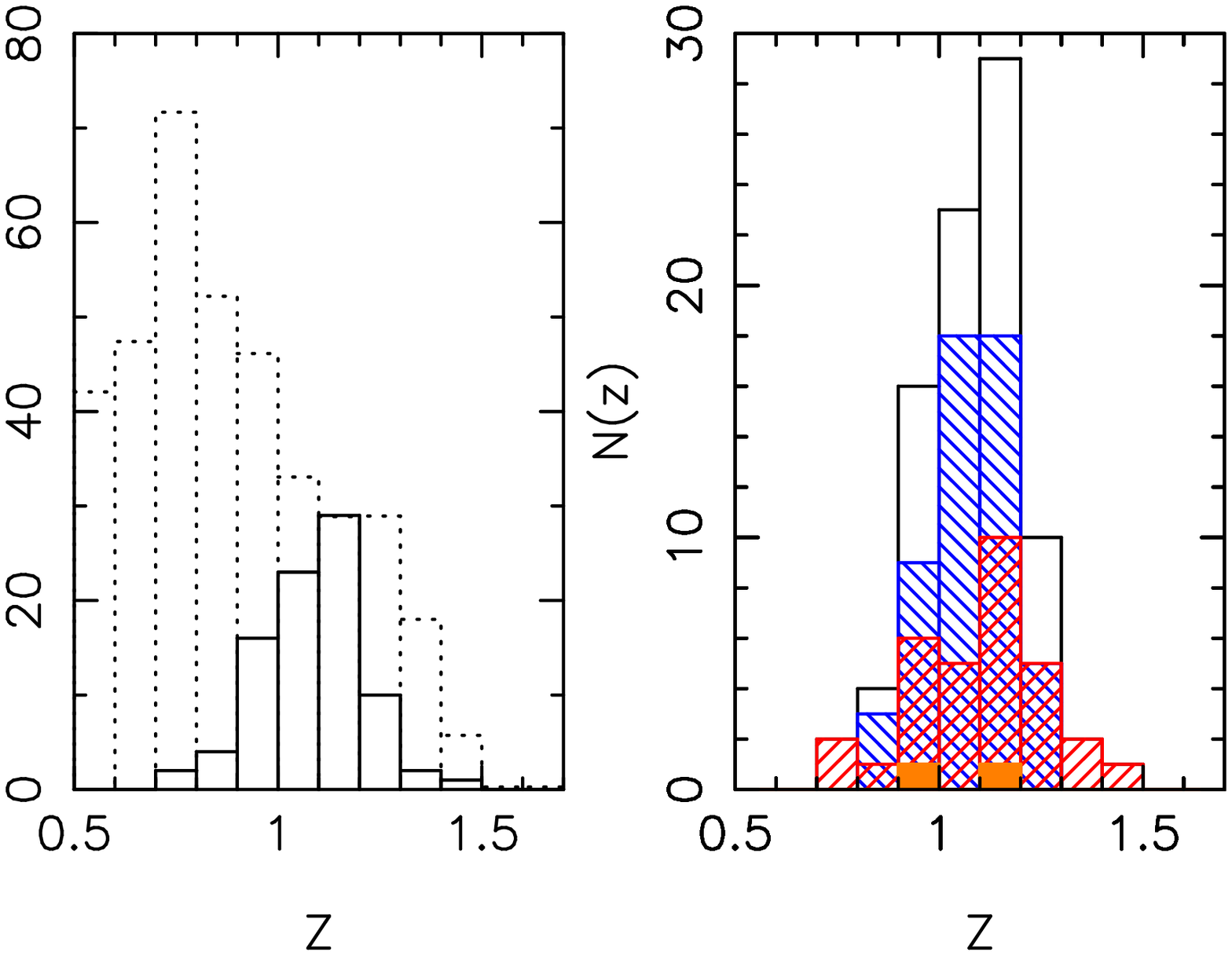}
\plotone{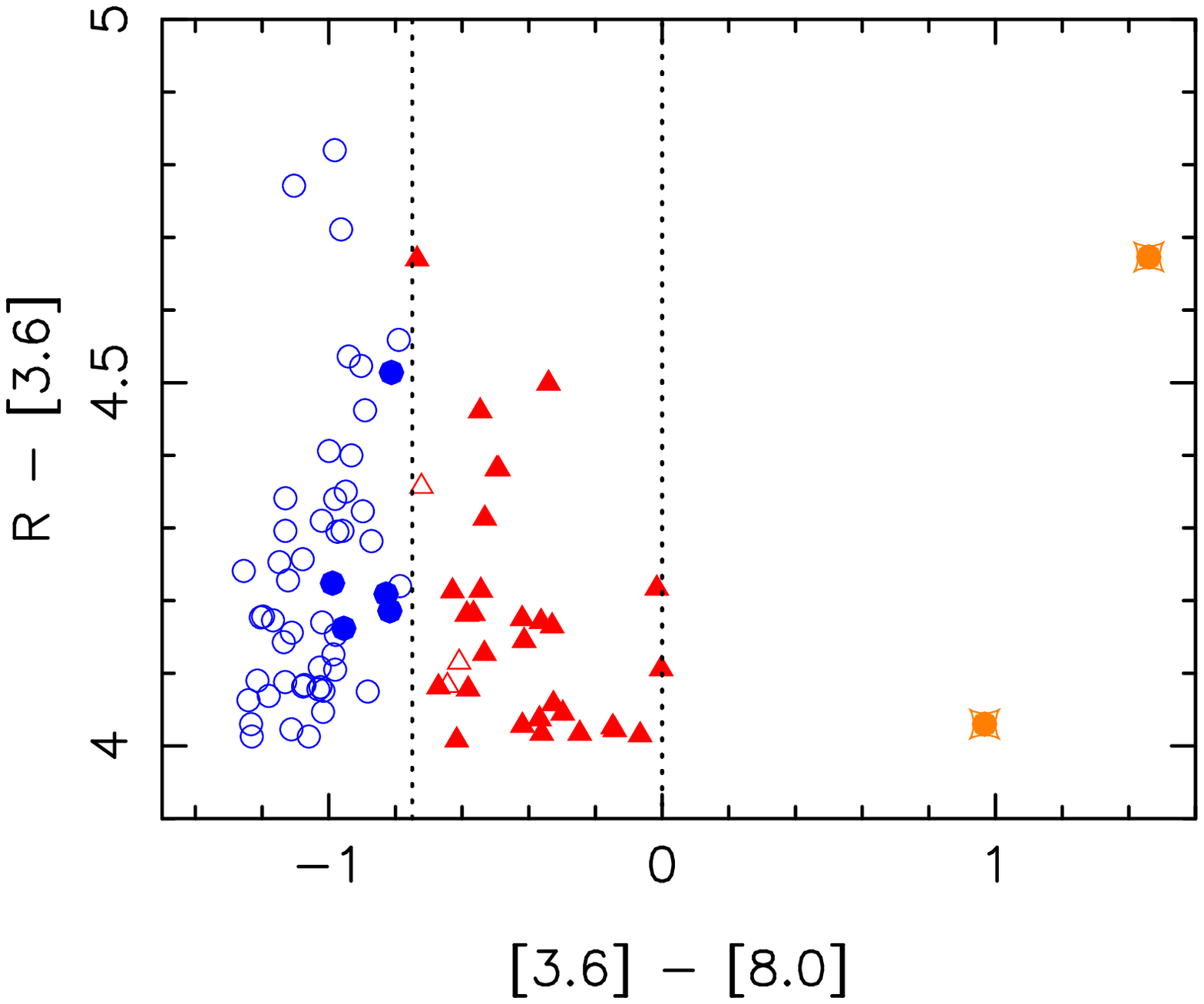}
\caption{
The solid black histogram in the far left panel shows the ERO  redshift distribution $N(z)$  (which peaks  for this sample of 87 EROs at $z=1.15$). 
The dotted black line illustrates the 
underlying DEEP2 selection function in the EGS field \emph{before} 
the ERO criterion is applied. The  dotted black line  shows the redshift distribution $N(z)$ of all 6023 
galaxies with good quality 
redshift determinations, scaled down by a factor of 15 for comparison with the ERO $N(z)$. 
The right panel (dashed black rectangle in  Figure~\ref{fig:R14}) shows our IRAC $[3.6] - [8.0]$ color division into 
early type (blue circles), dusty starburst (red triangles) and power-law (orange boxes) EROs.
Filled symbols indicate the 36 EROs with a $24\micron$ detection (19.1 AB, $5\sigma$, \S~\ref{sec:disc}). 
Also shown in the middle panel (and discussed in \S~\ref{sec:IRACcolorclass}), is the redshift distribution of the full sample (black) and the three
sub-classes (using the same blue, red and orange color scheme). 
\label{fig:hist}
}
\end{figure}

\begin{figure}
\epsscale{1.0}
\plotone{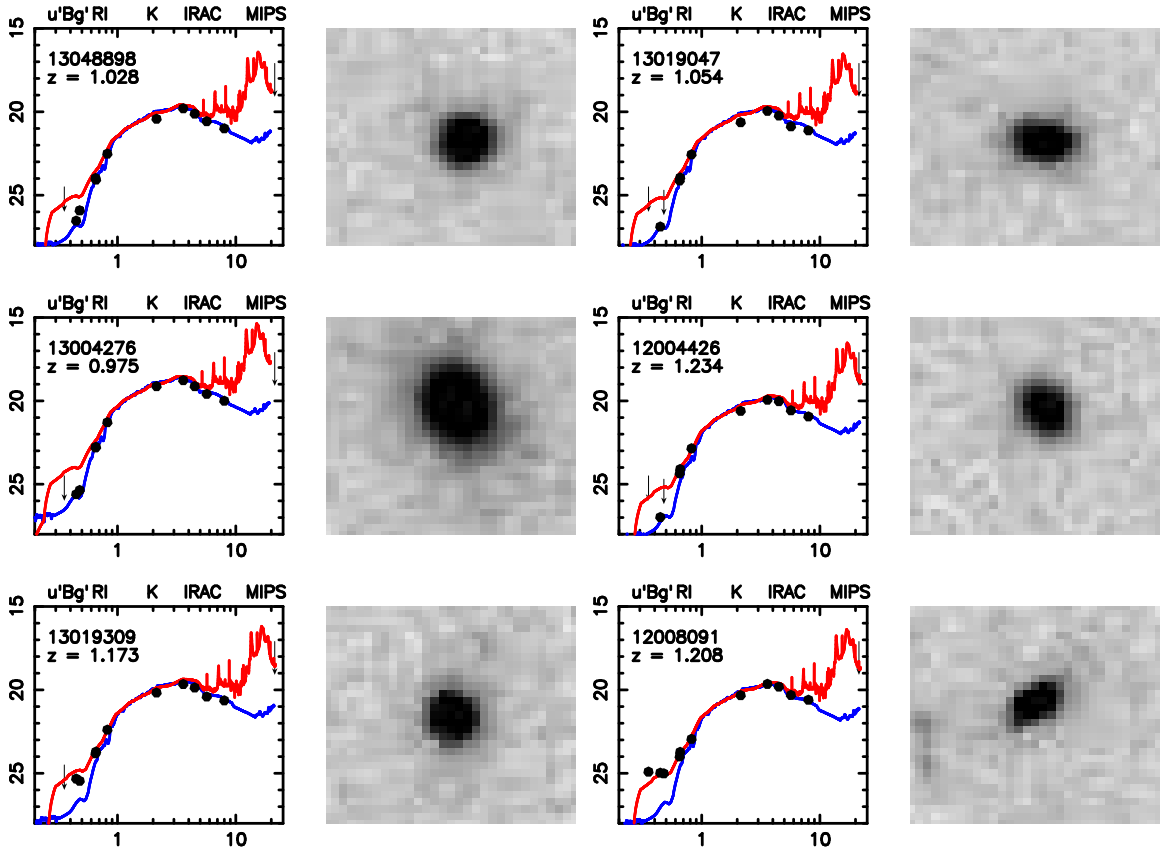}
\caption{
Examples of observed SEDs (left) and $3 \arcsec \times 3 \arcsec$ $I$-band ACS images (right) for 
six IRAC color-selected early-type EROs. A maximum of 
12 passbands ($u^{\prime}$, $B$, $g^{\prime}$, Subaru $R$, CFHT $R$, $I$, $K$
[3.6], [4.5] [5.8], [8.0] and [24.0]) are shown. The tip of the downward pointing arrow indicates upper limits in the case of non-detections. Each panel shows the DEEP2 ID number, and the spectroscopic redshift. Also shown
are a CWW E (blue) and M82 (red) template (\S~\ref{sec:color-color}) as they would appear
 at the redshift of each ERO. See  \S~\ref{sec:seds} discussion.
\label{fig:seds_E}
}
\notetoeditor{This figure is intended to comprise upper half of Plate 1}
\end{figure}

\begin{figure}
\plotone{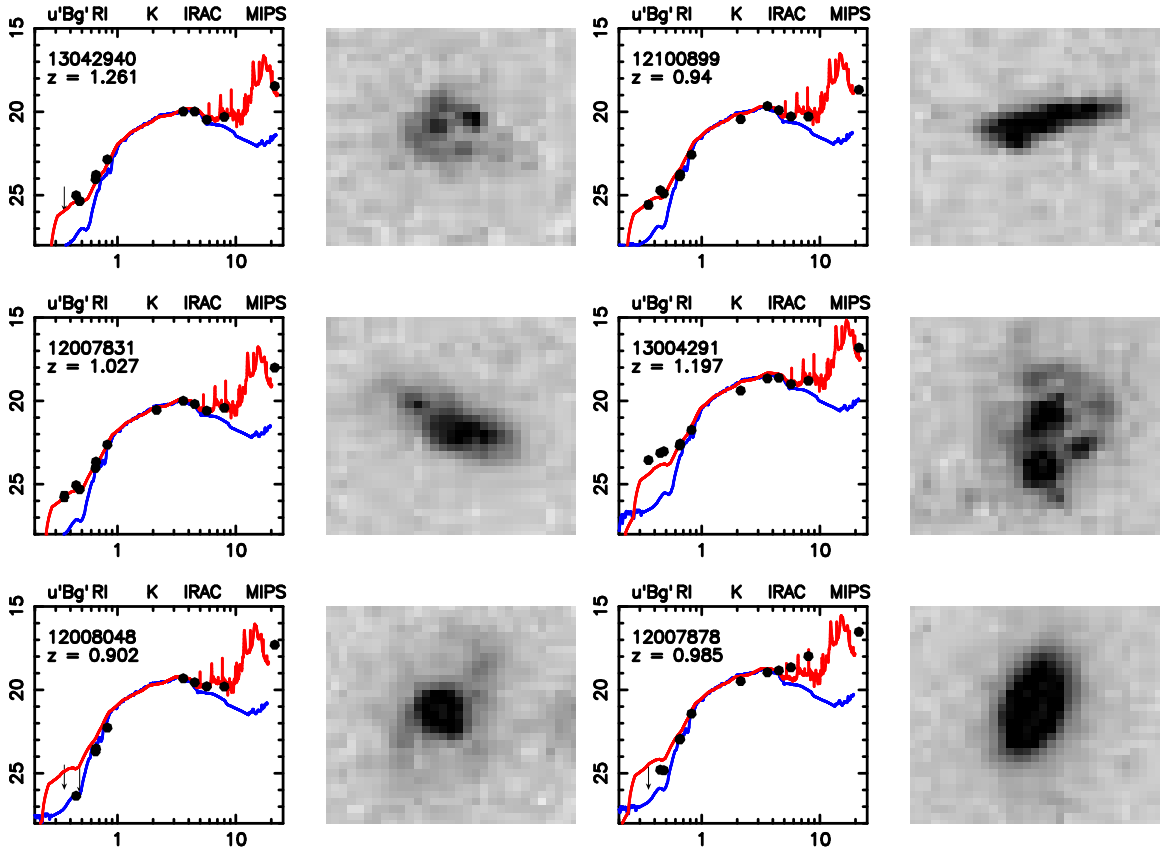}
\caption{
As for Figure~\ref{fig:seds_E}, but for five IRAC color-selected dusty starburst EROs 
and one power-law ERO (12007878).
The M82 dusty starburst template approximates the 13042940, 12100899, and  12007831 SEDs very well,
and that of 13004291 reasonably well.
In most cases the ACS images show clear evidence of interactions and mergers.  12008048 is 
a face-on dusty spiral galaxy.
\label{fig:seds_DSB}
}
\notetoeditor{This figure is intended to comprise lower half of Plate 1}
\end{figure}

\begin{figure}
\plotone{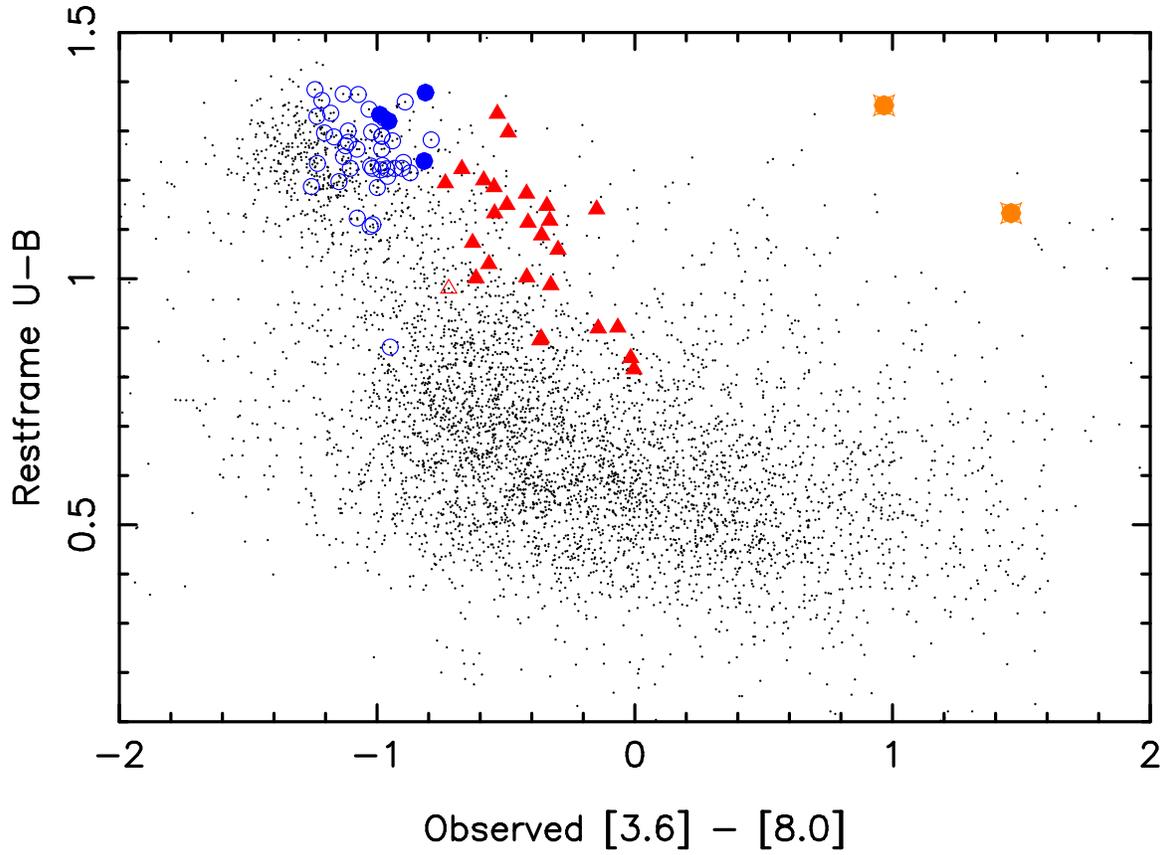}
\caption{
Restframe $U-B$  color versus observed $[3.6]-[8.0]$ color for all IRAC-detected galaxies with
good [8.0] \emph{and} good CFHT BRI photometry. Note: $(U-B)_{\rm Vega}$ = $(U-B)_{\rm AB}$ - 0.85. Blue circles indicate 
bulge-dominated early-type EROs. Red triangles indicate dusty starburst EROs.
Orange boxes indicate the two AGN.  Filled symbols indicate a $24\micron$ detection.
Is is possible that the $24\micron$ detections associated with
each of the early-type EROs may be spurious, and actually associated with  a close 
neighbor galaxy (see \S~\ref{sec:disc}). The early-type EROs generally tend to 
lie on the $U-B$ defined red-sequence and  the dusty starburst  EROs lie in the 
``green valley'' and at the top of the ``blue cloud''.
\label{fig:UB14}
}
\end{figure}

We showed that the three classes of ERO and their redshift distribution 
were broadly consistent with what one would expect from a simple color-color diagram (Figure~\ref{fig:R14}). 
We found 53 early-types, 32 dusty starbursts/LIRGs, and two obvious AGN.  Both of these AGN
would be classified as ULIRGs.

The selection of this particular sample of EROs was subject to the 
spectroscopic biases  discussed in  \S~\ref{sec:zdist}.
We next plan to extend this study to the  several thousand IRAC-selected EROs in the 
AEGIS \emph{without} spectroscopic redshifts.

\acknowledgements 

This work is based in part on observations made with the Spitzer
Space Telescope, which is operated by the Jet Propulsion Laboratory,
California Institute of Technology under a contract with NASA.
Support for this work was provided by NASA through an award issued by
JPL/Caltech. 
ALC is supported  
by NASA through Hubble Fellowship grant HF-01182.01-A.
We thank the referee, Matt Malkan, for constructive comments.
GW thanks Mark Lacy, Adam Muzzin, Jason Surace, Ian Smail and Mike Hudson for useful discussions,
and UC Berkeley and UC Santa Cruz for their hospitality.

\newpage


\clearpage

\clearpage

\clearpage

\clearpage

\clearpage

\end{document}